\documentclass[reprint,amsmath,amssymb,prb,superscriptaddress]{revtex4-2}
\usepackage{graphicx}
\usepackage{dcolumn}
\usepackage{bm}
\usepackage{epstopdf}
\usepackage{adjustbox}
\usepackage{float}
\usepackage{cancel}
\usepackage[normalem]{ulem}
\usepackage{xcolor}
\usepackage{amssymb}
\usepackage{amsmath}
\usepackage{graphicx}
\usepackage{epstopdf}
\usepackage{soul}
\usepackage[breaklinks=true,colorlinks=true,linkcolor=blue,urlcolor=blue,citecolor=blue]{hyperref}
\usepackage[utf8]{inputenc}
\usepackage[T1]{fontenc}
\usepackage{mathptmx}
\begin{document}
	
	\preprint{APS/123-QED}

	
	\title[mode = title]{Field-free ultrafast magnetization reversal of a nanodevice by a chirped current pulse via spin-orbit torque}
	
	\author{Y. D. Liu}
	\affiliation{Center for Spintronics and Quantum System, State Key Laboratory for Mechanical Behavior of Materials, School of Materials Science and Engineering, Xi’an Jiaotong University, Xi’an, Shaanxi, 710049, China}
	
	\author{M. T. Islam}
	\email[Correspondence email address:]{torikul@phy.ku.ac.bd}
	\affiliation{Center for Spintronics and Quantum System, State Key Laboratory for Mechanical Behavior of Materials, School of Materials Science and Engineering, Xi’an Jiaotong University, Xi’an, Shaanxi, 710049, China}
	
	\affiliation{Physics Discipline, Khulna University, Khulna 9208, Bangladesh} 
	
	\author{T. Min}
	\email[Correspondence email address:] {tai.min@xjtu.edu.cn}
	\affiliation{Center for Spintronics and Quantum System, State Key Laboratory for Mechanical Behavior of Materials, School of Materials Science and Engineering, Xi’an Jiaotong University, Xi’an, Shaanxi, 710049, China}
	
	\author{X. S. Wang}
	\affiliation{School of Physics and Electronics, Hunan University, Changsha 410082, China }
	
	\author{X. R. Wang}
	\affiliation{Physics Department, The Hong Kong University of Science and Technology, Clear Water Bay, Kowloon, Hong Kong}
	\affiliation{HKUST Shenzhen Research Institute, Shenzhen 518057, China}

	\begin{abstract}
We investigated the magnetization reversal of a perpendicularly magnetized nanodevice using a chirped current pulse (CCP) via spin-orbit torques (SOT). Our findings demonstrate that both the field-like (FL) and damping-like (DL) components of SOT in CCP can efficiently induce ultrafast magnetization reversal without any symmetry-breaking means. For a wide frequency range of the CCP, the minimal current density obtained is significantly smaller compared to the current density of conventional SOT-reversal. This ultrafast reversal is due to the CCP triggering enhanced energy absorption (emission) of the magnetization from (to) the FL- and DL-components of SOT before (after) crossing over the energy barrier. We also verified the robustness of the CCP-driven magnetization reversal at room temperature. Moreover, this strategy can be extended to switch the magnetic states of perpendicular synthetic antiferromagnetic (SAF) and ferrimagnetic (SFi) nanodevices. Therefore, these studies enrich the basic understanding of field-free SOT-reversal and provide a novel way to realize ultrafast SOT-MRAM devices with various free layer designs: ferromagnetic, SAF, and SFi.
\end{abstract}

	\maketitle
	
	\section{Introduction}
In recent decades, spin transfer torque (STT) has been the primary means of manipulating the magnetic states of ferromagnetic (FM) and antiferromagnetic (AFM) materials using spin-polarized current \cite{slonczewski1996, tsoi1998, Katine2000, Waintal2000, sun2000a, stiles2002, bazaliys2004, koch2004, wetzels2006, Ikeda2007, manchon2008, miron2010, Wadley2016, Fukami2016, Wang2013V, miron2011perpendicular,Liu2012, Endoh2018,elezn2014}. However, spin-orbit torque (SOT) has emerged as an efficient alternative means of controlling and manipulating the states of FM/AFM materials using electric current \cite{Gambardella2011, miron2011perpendicular, Liu20121, parkin1991systematic, Fukami2016}. SOT-switching offers promising advantages over STT-switching, such as high-speed and energy-efficient switching without the need for the switching current to pass through a tunnel barrier, which is necessary for magnetic random access memory (MRAM) \cite{miron2011perpendicular,Liu2012,Yu2014,Fukami2016}.

Typically, to obtain SOT-switching, a current is applied in a heavy-metal (HM)/topological insulator(TI) layer on which a FM/AFM free layer is placed \cite{Liu2012,Fukami20161,Manchon2019, Ryu2020}. The transverse spin currents generated in the HM/TI layer due to the spin-Hall effect \cite{Sinova2015} or Rashba effect \cite{Manchon2019} pass through the free layer positioned on top of the HM layer and induce SOT, consisting of mutually orthogonal field-like (FL) and Slonczewski (damping-like, DL) torques, on the free layer magnetization. However, in most conventional SOT-reversal studies, only the DL-torque component is utilized \cite{Kim2012,Fukami20161,Chen2017}, which results in a high minimal current density required for the desired reversal rate. Additionally, conventional SOT-switching requires an external field or other means (e.g., lateral asymmetric introduced by the thickness variation of layers and an exchange bias field created by antiferromagnetic) to ensure deterministic switching \cite{Yu20142,Yu2014,Chen2018}.
Recent experimental works have shown that the FL-torque ($\tau_\text{FL}$) can be significantly larger than the DL-torque ($\tau_\text{DL}$) (i.e., $|\tau_\text{FL}/\tau_{DL}|\approx4$) in certain nanostructure (e.g.,Ta/CoFeB/Mgo), and the substantial FL-torque can induce magnetization reversal \cite{JKim2012, Garello2013, Avci2014, Qiu2014, Akyol2015, Yoon2017, Wang2012}.  For dominant FL-torque reversal, a properly designed chirped current pulse, analogous to the chirped magnetic field pulse-induced magnetization switching \cite{thirion2003, rivkin2006, Woltersdorf2007, sun2006_97, sunz2006_73, sun2006magn, barros2011, barrose2013, cai2013, islam2018, islam2020}, can be an efficient alternative to drive ultrafast field-free magnetization reversal. However, a recent study \cite{Zhang2018} proposed a theoretical current pulse based on the constraint of the shortest time, which is difficult to realize practically to obtain the ideal balance of the FL- and DL-torques. Another recent study \cite{Vlasov2022} identified an analytical in-plane current pulse to switch magnetization by balancing the FL- and DL-components of SOT during the switching process. However, the main parameters of this current pulse change with time, making it complex to realize in practice. Therefore, it is still desirable to find a simpler setup with a clear physical mechanism for field-free, fast, and energy-efficient SOT-reversal from the viewpoint of fundamental understanding and applications.

In this study, we investigate magnetization reversal of a perpendicularly magnetized nanodevice by a chirped current pulse (CCP) via SOT. Our findings demonstrate that both FL- and DL-components of SOT in CCP can be efficiently utilized to induce ultrafast magnetization reversal without any symmetry-breaking means. For a wide frequency range of the CCP, the obtained minimal current density is significantly smaller compared to the current density of conventional SOT-reversal and the theoretical limit \cite{Zhang2018}. This ultrafast reversal is due to the mechanism that the CCP triggers the enhanced energy absorption (emission) of the magnetization from (to) the FL- and DL-components of SOT before (after) crossing over the energy barrier. We also verify the robustness of the CCP-driven magnetization reversal at room temperature. Moreover, this strategy can be extended to switch the magnetic states of perpendicular synthetic antiferromagnetic (SAF) and ferrimagnetic (SFi) nanodevices. Therefore, these studies enrich the basic understanding of field-free SOT-reversal as well as provide a novel way to realize the ultrafast SOT-MRAM device with various free layer designs: ferromagnetic, SAF, and SFi.

	\begin{figure}
		\includegraphics[width=85mm,height=45mm]{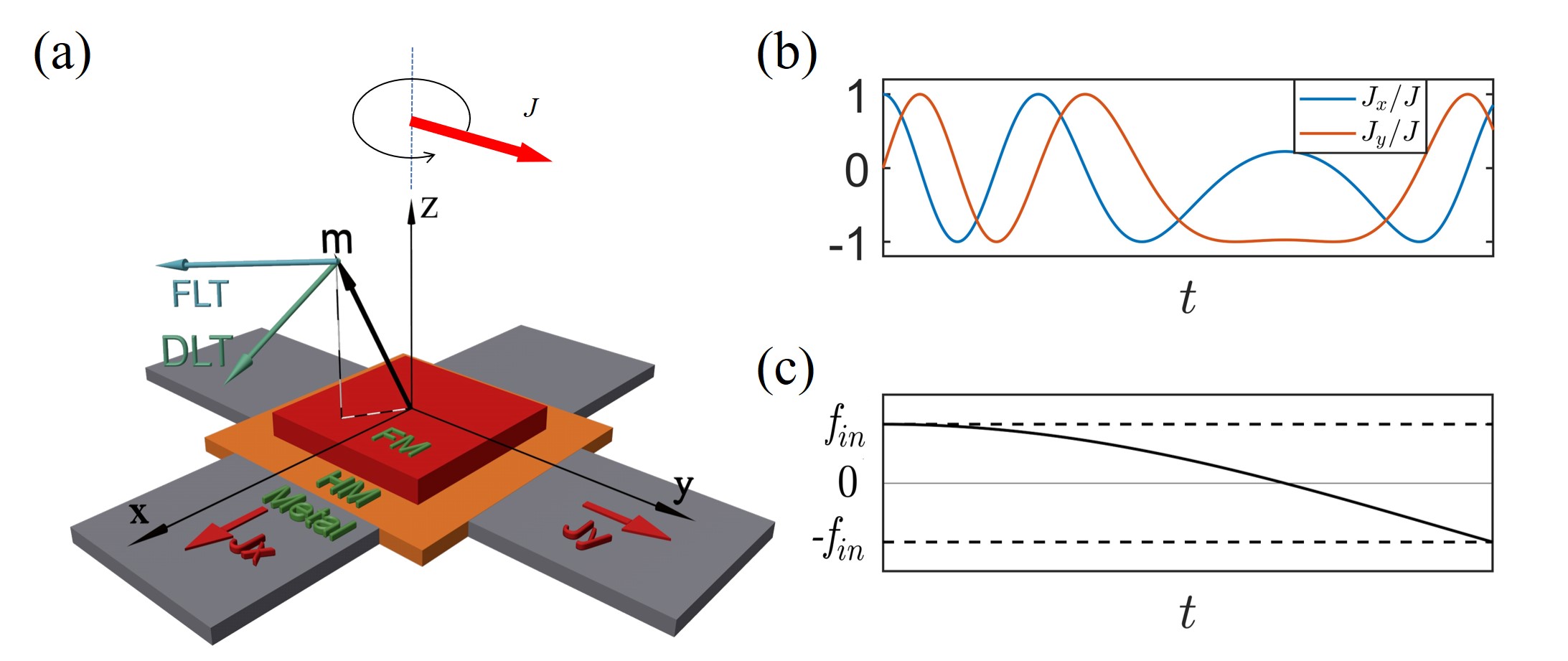}
		\caption{\label{Fig1} (a) Schematic representation of the nanodevice with SOT components on  \(\mathbf{m}\) which is perpendicular to \(xy\)-plane initially. Two components of $\mathbf{J}$ are applied along $+x$ and $+y$ directions in the HM. (b) Temporal change of the orthogonal components of the normalized electric current which flow in the heavy metal layer.  (c) The frequency sweeping profile (\(+f_{\text{in}}\) to  \(-f_{\text{in}}\)) of CCP with time.}
	\end{figure}
	
	\section{Model and methodology}
	The nanodevice under consideration consists of a ferromagnetic free layer and a heavy metal (HM) layer, as shown in Fig. \ref{Fig1}(a). The free layer's magnetization $\mathbf{m}$ is perpendicular to the $xy$-plane, with two energy minima along the +$z$ and -$z$ directions due to easy-axis anisotropy. To reverse the magnetization from one stable state to another, a time-dependent rotating electric current pulse $\mathbf{J} = J\left[\cos\phi_\text{j}(t)\hat{\mathbf{x}}+\sin\phi_\text{j}(t)\hat{\mathbf{y}}\right]$ is applied, with its two components along the +$x$ and +$y$ directions in the HM layer. The magnitude of $\mathbf{J}$ remains constant, and the phase angle $\phi_\text{j}(t)$ between $\mathbf{J}$ and $\hat{x}$ varies with time. As the current passes through the HM layer, a transverse spin current is generated due to the spin Hall or Rashba effects, which exerts an effective torque on the magnetization \cite{Hirsch1999,MacNeill2016,Humphries2017}. This torque consists of field-like (FL) and damping-like (DL) components, as shown in Fig. \ref{Fig1}(a),
	\begin{equation}
		\pmb{\tau} = -\tau_{\text{DL}}\mathbf{m}\times(\mathbf{m}\times \mathbf{p})-\tau_{\text{FL}}(\mathbf{m}\times \mathbf{p}),
		\label{SOT}
	\end{equation}
	$\tau_{\text{DL}} = (\hbar/2e)(J/M_{\text{s}}d)c^{\parallel}$ and $\tau_{\text{FL}} = (\hbar/2e)(J/M_{\text{s}}d)c^{\perp}$ magnitudes (in the unit of magnetic field) of the DL and FL-torques , respectively, $d$ is the thickness of the ferromagnetic layer, $c^{\parallel}$ and  $c^{\perp}$ are the efficiencies of DL and FL torque respectively. In this study, we use relative strength $\xi=-3$ which is the ratio of the $\tau_{\text{FL}}$ to $\tau_{\text{DL}}$.   $\mathbf{p}=\hat{J}\times\hat{z}$ is the spin-polarization direction, where $\hat{J}$ is the electric current direction.
	
In presence of the electric current density, $\mathbf{J}$, the magnetization dynamics is described by the explicit Landau-Lifshitz-Gilbert equation \cite{gilbert2004}
	\begin{align}
		\dfrac{1+\alpha^2}{\gamma}\dfrac{d\mathbf{m}}{dt}=&-\mathbf{m}\times\mathbf{H}_\text{eff}-\alpha\mathbf{m}\times(\mathbf{m}\times\mathbf{H}_\text{eff})\notag \\ 
		&-\tau_{\text{DL}}[(1+\xi\alpha)\mathbf{m}\times(\mathbf{m}\times\mathbf{p})+(\xi-\alpha)(\mathbf{m}\times\mathbf{p})], 
		\label{llg}
	\end{align}
	where  $\alpha$ and $\gamma$ are the Gilbert damping constant and gyromagnetic ratio respectively.  The effective field $\mathbf{H}_\text{eff}$ comes from the exchange field $\dfrac{2A}{\mu_0M_\text{s}}\nabla^2 \mathbf{m}$, and the easy-axis anisotropy field $H_k$ along  $\hat{z}$ direction. 
	\begin{figure*}
		\includegraphics[width=130mm,height=40mm]{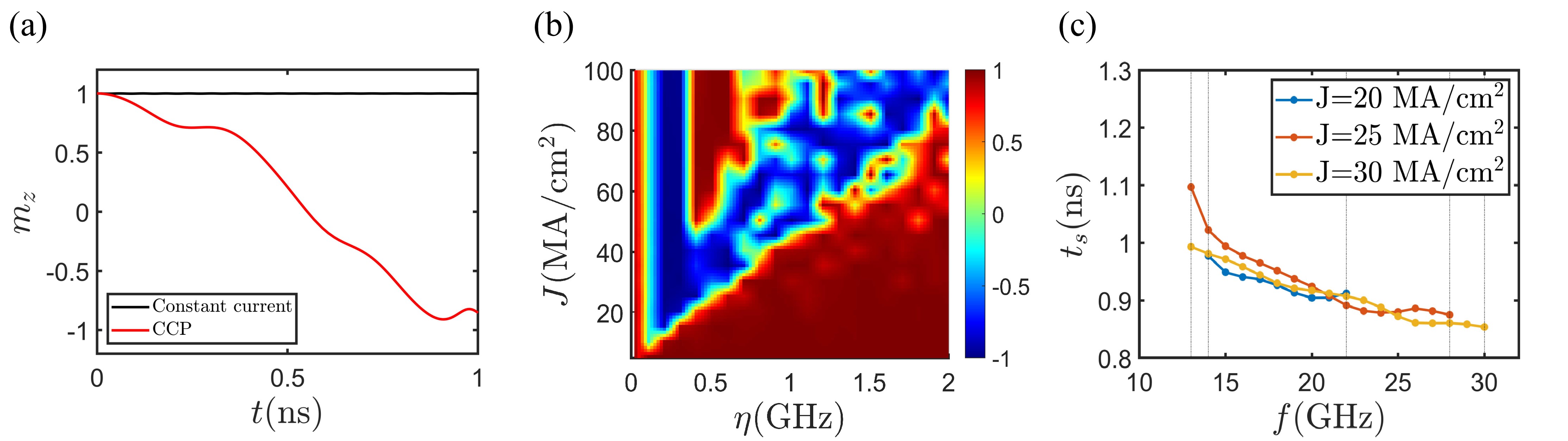}	
		\caption{\label{Fig2} (a) Red line shows the temporal change of \(m_z\)  under the CCP with \({f_\text{in}}\) = 16.8 GHz, \(J = 20\) MA/cm\(^2\) and \(\eta=\)  0.22 GHz and black line represents the \(\textbf{m}\) stays initial state for 1D  current of   \(J = 20\) MA/cm\(^2\). (b) For fixed \({f_\text{in}}\) = 16.8 GHz,  the phase diagram (\(\textbf{J}-\eta\)) of magnetization switching in terms of current amplitude \({J}\)  and \(\eta\) of CCP. (c) For fixed \(\eta=\) 0.22 GHz, the switching time, \(t_s\) as a function of \({f_\text{in}}\) for different \(J\).}
	\end{figure*}
	From the Eq. \eqref{llg}, it is noted that the DL and FL-torque fields interact with the magnetization in two different means, specifically, the DL torque compels the magnetization along $y-$axis while the field from FL-torque generates  the precession of magnetization around $z-$axis. 
	However,  there are the studies \cite {Legrand2015,Yoon2017} which have taken into account the FL-torque as a dominant driving force as in some heavy metal, the FL-torque is significant.  
		
	The spin-orbit torque is a non-conservative force which works on the magnetization. Under this non-conservative force, the energy changing rate (with dimension) of the system is presented as (refers to Appendix for detail derivation)
	\begin{align}
		\dot\epsilon =& -\frac{\alpha \gamma \mu_0 M_S}{1+\alpha^2}[|\mathbf{H}_{\text{eff}}|^2 - (\mathbf{m} \cdot \mathbf{H}_{\text{eff}})^2]+ \dot\epsilon_\text{SOT}.
		\label{totaleng}
	\end{align}
	
The first term can be zero or negative as it is proportional to $-\alpha$, and $|\mathbf{H}_{\text{eff}}|$ is always no larger than $|\mathbf{m} \cdot \mathbf{H}_{\text{eff}}|$. This term reflects energy dissipation due to the Gilbert damping.  The second term, \(\dot\epsilon_\text{SOT}=- \frac{\gamma \mu_0 M_S a_j}{1+\alpha^2} \mathbf{H}_{\text{eff}}\cdot [(1+\xi \alpha) \mathbf{m} \times (\mathbf{m} \times \mathbf{p}) + (\xi - \alpha)(\mathbf{m} \times \mathbf{p})]\), can be either positive or negative, contributed by the effective SOT which is generated by the time dependent spin current. Alternatively, the equivalent fields of the FL- and DL-torque can either supply the energy to the magnetization or extract the energy (by negative work done) from the magnetization, depending on the angle between the instantaneous magnetization direction $\mathbf{m}$ and spin polarization direction $\mathbf{p}$ \cite{islam2018, islam2021}. 

Without any external forces, the easy-axis anisotropy allows the magnetization $\mathbf{m}$ of the free layer to stay in two equilibrium states/energy minima,  $\mathbf{m}\parallel \hat{\mathbf{z}}$ and $\mathbf{m}\parallel -\hat{\mathbf{z}}$. The main objective of deterministic switching is to move the $\mathbf{m}$  from one stable state to another. Across the path of switching, $\mathbf{m}$ requires to climb the energy barrier and  cross  the equator of energy barrier at $\mathbf{m}_{z} = 0$ and then, because of the Gilbert damping, $\mathbf{m}$ goes down to another stable state.  So the ultrafast and energy efficient reversal might be obtain if the $\mathbf{m}$ resonantly absorbs energy from external source to reach at the equator and then releases the energy  by the negative work-done. In the context of magnetic field driven, such fast and the energy efficient reversal has been demonstrated by the linear down chirped/cosine chirped microwave pulse \cite{islam2018, islam2021}  through the physics of the  microwave energy absorption (emission) by (from) the  $\mathbf{m}$ before (after) the equator of energy barrier. To avail the similar mechanism, we employ the CCP (cosine) to utilize FL and DL-torque fields effectively. The CCP can be recast as the form 
	\begin{equation}
		\mathbf{J}(t)=J\left[\cos\phi_\text{j}(t)\hat{\mathbf{x}}+\sin\phi_\text{j}(t)\hat{\mathbf{y}}\right]\label{app},
		\end {equation} 
		where $J$ is the magnitude of the electric current and $\phi_\text{j}(t)$ is the time-dependent phase angle.   $\phi_\text{j}(t)$ is considered as $2 \pi f_{\text{in}} \cos \left(2 \pi \eta t \right) t$, where $\eta$ (in GHz unit) is the controlling parameter. The frequency of CCP seeps  from initial $f_{\text{in}}$ to final $-f_{\text{in}}$ with the seeping rate  $k(t)$ (in units of $\text{ns}^{-2}$) while $k(t)$ takes the form $ - f_{\text{in}} \left[\left({4 \pi} \eta \right) \sin\left({2 \pi \eta t} \right) + \left({2 \pi}\eta \right)^2 t \cos\left({2 \pi \eta t} \right) \right]$. The time- dependent frequency is represented as $f(t) = \dfrac{1}{2\pi} \dfrac{d\phi_\text{j}}{dt} = f_{\text{in}} \left[\cos \left(2 \pi \eta t \right) - \left(2 \pi \eta t\right) \sin \left(2 \pi \eta t\right) \right]$ which is shown in Fig. \ref{Fig1}(c).
	   If the pulse duration (time requires to sweep $+f_{\text{in}}$ to final $-f_{\text{in}}$) is $t_p$, by solving the condition $\cos(2\pi \eta t_p)-(2\pi \eta t_p)\sin(2\pi \eta t_p)=-1$, we get a relation between $t_p$ and \(\eta\) as follows,
		\begin{equation}
			t_p=\frac{1.307}{2\pi \eta}. 
			\label{tp}
		\end{equation}

		For this study, we have used the micromagnetic simulation package, MuMax3 \cite{vansteenkiste2014} to solve the LLG equation numerically. From the viewpoint of experimental realization, we choose the material parameters: saturation magnetization,  $M_\text{s}=10^6$ A/m, perpendicular anisotropy $K_\text{u} = 8\times10^5$ J/m$^3$ or $h_\text{ani}=1.6$ T,  $\gamma = 1.76\times 10^{11}$ rad/T/s, exchange constant \(A_\text{ex} = 1.5\times 10^{-11}\)  J/m, and Gilbert damping $\alpha$= 0.01.  The system we focused on has a dimension of $20\times 20 \times 1.2$ nm$^3$, and it was discretized with a mesh size of $1 \times 1 \times 1.2$ nm$^3$.

		\begin{figure}
			\includegraphics[width=85mm,height=70mm]{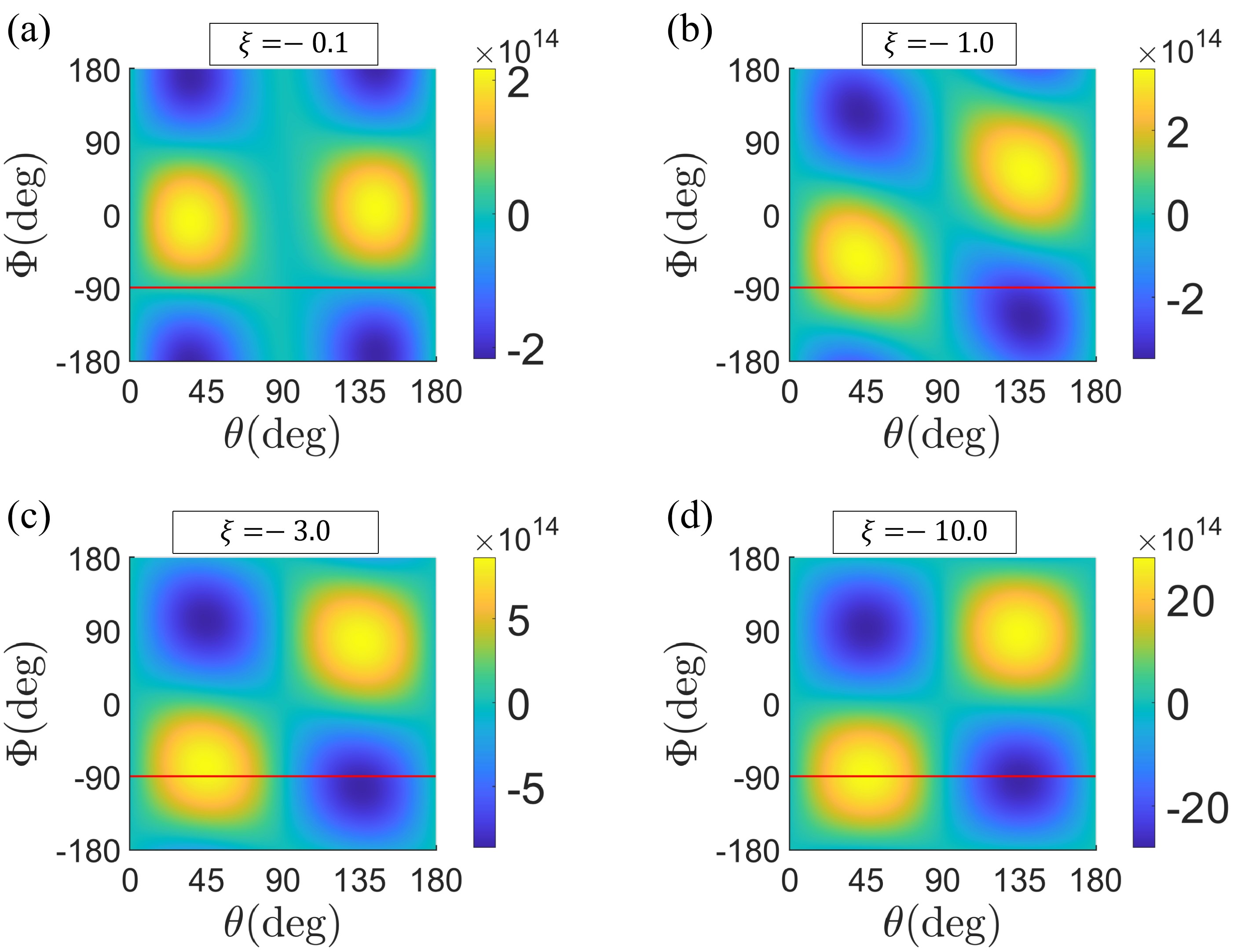}
			\caption{\label{Eng1}  Phase diagrams of the rate of energy change \(\dot\epsilon\)   in terms of $\theta$ and $\Phi$ for different  $\xi (\tau_{\text{FL}}/\tau_{\text{FL}})$ (a) -0.1, (b) -1, (c) -3 and (d) -10. The color bar represents \(\dot\epsilon\) while yellow (blue) color indicates \(+\dot\epsilon\) (-\(\dot\epsilon\)).   }
		\end{figure}

		\section{Numerical Results}

		Initially the magnetization $\mathbf{m}$  stays in one equilibrium state ($m_\text{z} = 1$) and its  resonant frequency, based on the system configuration and the uniaxial anisotropy ($h_\text{ani} $), is determined as $\dfrac{\gamma}{2\pi} \left[h_\text{ani} - \mu_0 (N_z-N_x)M_\text{s}\right] = 16.8$ GHz. To induce the magnetization switching, we apply the chirped current pulse (CCP) to the heavy metal (HM) layer by choosing the initial frequency of the CCP as  $f_\text{in} = 16.8$ GHz  so that it closely matches the magnetization precession frequency. Then we investigate the CCP driven switching with different values of the current $J$ and  $\eta$. Interestingly, with the minimal $J = 20$ MA/cm$^2$ and the optimal $\eta = $0.22 GHz , the CCP  is capable of inducing the fast and efficient deterministic switching (red line) which is shown as in the Fig. \ref{Fig2}(a). For fixed $f_\text{in} = 16.8$ GHz and the switching time window $t_s = $1.5 ns,  Fig. \ref{Fig2}(b) depicts phase diagram ($J-\eta$ ) in terms of $J$ and $\eta$ in which blue (red) color indicates the switched, $m_\text{z} =-1$ (not switched, $m_\text{z} = 1$) state. It is found that the critical switching density linearly increases with the increase of $\eta$. Later on, for different $J$ (20, 25 and 30 MA/cm$^2$)  of CCP with $\eta = $0.22 GHz, we investigate the magnetization switching by tuning  $f_\text{in}$ presented in the Fig. \ref{Fig2}(c). It is found, for each current, there is a range of $f_\text{in}$ (indicated by vertical lines) around $16.8$ GHz for which  the fast reversal is obtained. So, there is a great flexibility of choosing the parameters of the CCP which broadens the path of experimental realization. The pulse width of CCP is $t_p=0.92$ ns , refers to the Eq. \eqref{tp}, which is close to switching time 0.94 ns and for deterministic switching, the pulse width does not require to be precised. In addition, this strategy does not require the external field as like the conventional SOT switching which shows the remarkable advantage.

		To compare with the conventional switching, we apply 1D current along $y$ direction with same magnitude but the  $\mathbf{m}$  does not switch rather it stays in the initial state (black line) as shown in the Fig. \ref{Fig2}(a). Then we increase the current $J$ and for  $J=200$ MA/cm$^2$ (which is ten times larger then the amplitude of the CCP),  $\mathbf{m}$ reverses but its switched back (not shown) i.e., to obtain deterministic switching, the external field is required. To get a better sense about our strategy, we compare the obtained minimal $J$ to the theoretical limit of the current density  refers to the study \cite {Zhang2018} and found that the CCP with the minimal  $J=20$ MA/cm$^2$,  leads the fast switching, which is close to the theoretical limit, $J=7$ MA/cm$^2$ obtained for the same parameters. Although  $J=20$ MA/cm$^2$ is required  to switch perpendicular magnetization of high anisotropy field $h_\text{ani}=1.6$ T. So it is expected that the minimal $J$  would be reduced to the practically feasible range for the material of comparatively lower anisotropy field. 
		
		To elucidate the physical mechanism of the CCP driven fast reversal, we express the second term of the \eqref{totaleng} in the following form
		\begin{align}
			\dot\epsilon_\text{SOT}= &-\frac{\alpha \gamma \mu_0 M_S}{1+\alpha^2}(M_s N_{zz}+\frac{2K_u}{\mu_0 M_s})\cos\theta\sin\theta \notag \\
			&[(1+\xi\alpha)\cos\theta\cos\Phi(t)+(\xi-\alpha)\sin\Phi(t)],
			\label{soteng}
		\end{align}
		where \(\theta\) and \(\phi_\text{m}\) are a polar and azimuthal angles of \(\mathbf{m}\). \(\Phi(t)=\phi_\text{p}(t)-\phi_\text{m}(t)\) is the relative phase angle between the in-plane components of \(\mathbf{m}\) and spin polarization direction \(\mathbf{p}\) while $\phi_\text{p}(t)$ is phase angle of \(\mathbf{p}\). For the fixed material parameters, the rate of energy contribution from SOT components, \(\dot\epsilon\) is found to be dependent on  $\theta$ and the $\Phi(t)$. For different values of $\xi (\tau_{\text{FL}}/\tau_{\text{FL}})$, we analytically determine the \(\dot\epsilon\) as a function of $\theta$, varies from 0$^{\circ}$ (initial state) to 180$^{\circ}$ (reversed state) and  $\Phi(t)$,  varies from -180$^{\circ}$  to 180$^{\circ}$ and thus draw the phase diagrams of \(\dot\epsilon\) in terms of $\theta$ and $\Phi$ as shown in the Fig. \ref{Eng1}. Particularly, the phase diagrams of \(\dot\epsilon\) for  $\xi=-0.1$, $-1.0$, $-3.0$ and  $-10.0$ are depicted in Figs. \ref{Eng1}(a)-(d) respectively in which the color bars represent the energy changing rate \(\pm\dot\epsilon\). Yellow and blue colors indicate \(+\dot\epsilon_\text{SOT}\) (energy supplies to \(\mathbf{m}\) by the SOT components) and \(-\dot\epsilon_\text{SOT}\) (energy extracts from \(\mathbf{m}\) by the SOT components) respectively. For the dominant DL-torque, i.e., $\xi=-0.1$, yellow color in Fig. \ref{Eng1}(a) represents that the energy absorption is occurred at $\Phi=0$ before and after the equator which will not induce the deterministic switching. Then if we include the FL-torque component, that is, for  $\xi=-1.0$ (for equal contribution of SOT components), the energy absorption (emission) would be at smaller (larger) than  $\Phi=-90^{\circ}$ as indicated by horizontal red line in Fig.  \ref{Eng1}(b). Of course, it will be difficult to maintain two different $\Phi(t)$ before and after the equator.  If we further increase the FL-SOT component, i.e., for $\xi=-3.0$ (dominant FL-torque) as in Fig. \ref{Eng1}(c), the maximum energy absorption (emission) before (after) the equator would be occurred at around $\Phi(t)=-90^{\circ}$ as indicated by horizontal rel line. So, it is predicted that  $\Phi(t)$ is required to maintain  around -90$^{\circ}$ such that the energy absorption rate, \(+\dot\epsilon_\text{SOT}\) (the energy emission rate\(-\dot\epsilon_\text{SOT}\)) before (after) crossing the equator (\(\theta=90^{\circ}\)) becomes maximum which will lead fast and an energy-efficient switching. Finally, for  $\xi=-10.0$ (almost with FL-torque) as in Fig. \ref{Eng1}(d), the maximum energy absorption (emission) is occurred at $\Phi=-90^{\circ}$ which is similar prediction to the studies of microwave chirped pulse \cite{islam2018, islam2021}.

\begin{figure}[b]
			\includegraphics[width=85mm,height=45mm]{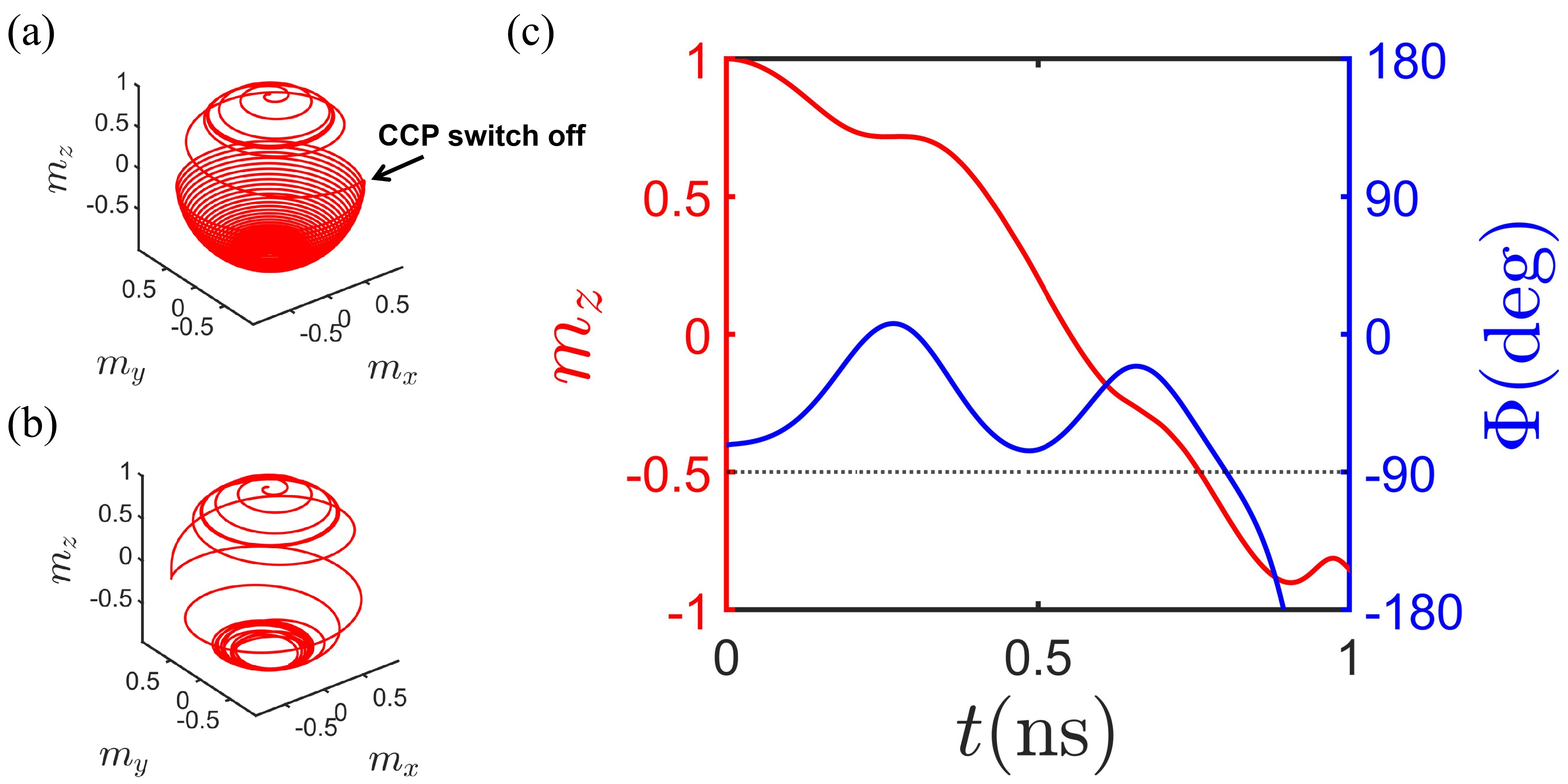}
			\caption{\label{Fig3} (a) The trajectory of magnetization if the CCP is switched off at \(m_\text{z}=-0.008\). (b) The trajectory of magnetization for the full CCP pulse, i.e., the CCP is applied until \(m_\text{z}\) goes to -1. (c) Red line depicts the temporal change of \(m_\text{z}\) and blue line shows the change of relative-angle \(\Phi\) with time while horizontal dotted-line indicates \(\Phi=-90^{{\circ}}\). }
		\end{figure}

		To understand the mechanism, one can look at the magnetization dynamics under CCP in detail. Fig. \ref{Fig3}(a) shows the switching trajectory in which the CCP is switched off at ${m}_z=-0.008$, i.e., just after crossing the equator.  $\mathbf{m}$  goes swiftly from \(m_\text{z}=1\) to \(m_\text{z}=-0.008\), after that,   $\mathbf{m}$ changes rotating direction and  goes to the ground state slowly as it requires to undergo many turns because of the damping coefficient. However, for the complete CCP, Fig. \ref{Fig3}(b) shows the magnetization switching trajectory and found that the $\mathbf{m}$ rotates anticlockwise (clockwise) before (after) crossing over the equator. Note that, after passing the equator, $\mathbf{m}$ rotates only a few turns and thus goes to the ground state swiftly. Compare to the former case (CCP is switched off at ${m}_z=-0.008$), it is evident that, after passing the equator, the CCP extracts the energy from the $\mathbf{m}$ which leads fast relaxation to the ground sate. Thus, the underlying mechanism is that $\mathbf{m}$ resonantly absorbs (emits) the energy before (after) crossing the equator from (to) the fields of DL- and FL-torque components.
			
\begin{figure}
			\includegraphics[width=85mm,height=60mm]{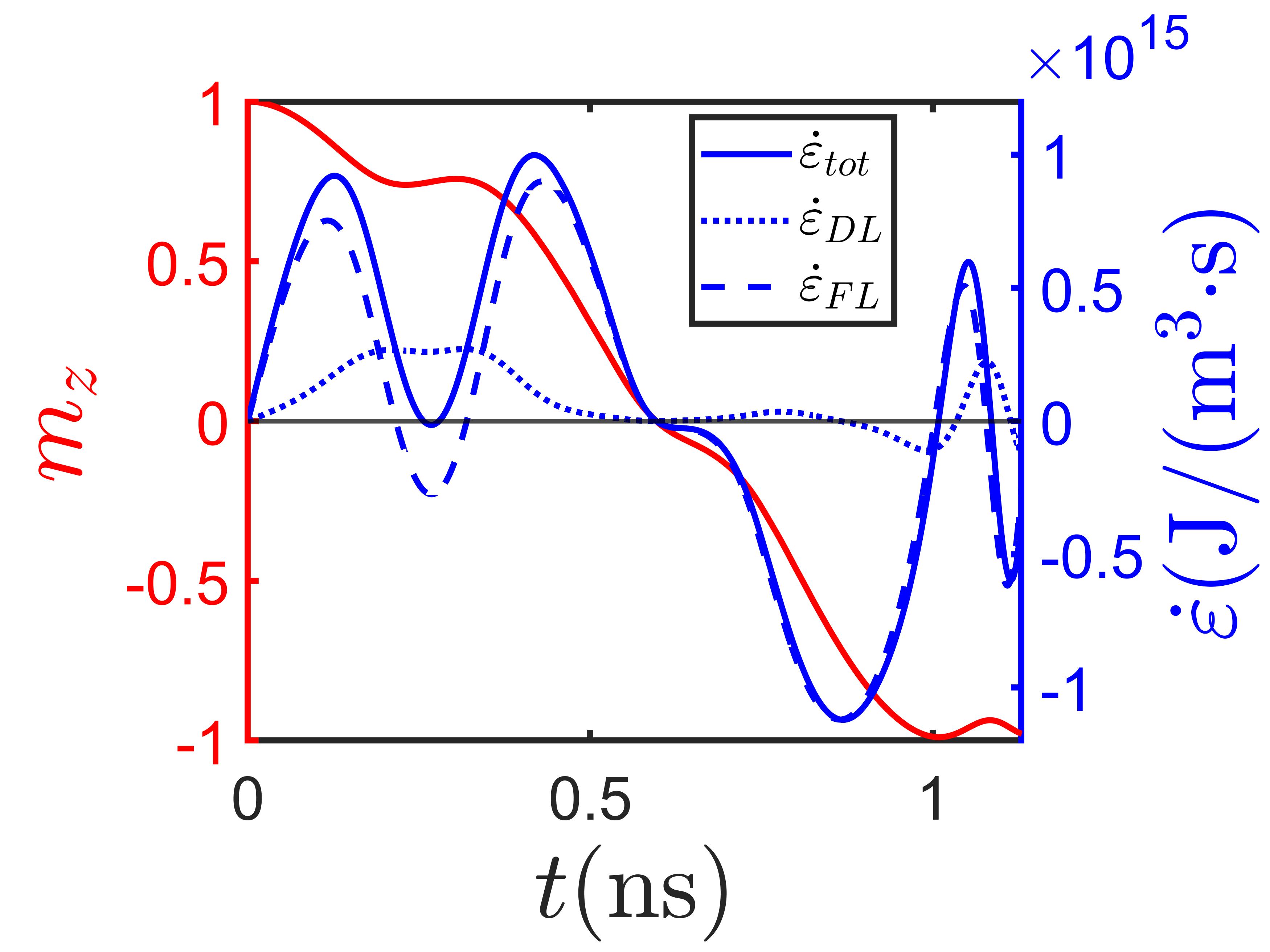}
			\caption{\label{Fig4} Red line shows change of \(m_\text{z}\) with time and plot of the energy changing rate of \(\mathbf{m}\) which contributed by the FL-torque (dashed blue line), DL-torque (dotted blue  line). Blue line shows the total energy change rate of the \(\mathbf{m}\) contributed by both SOT components.}
		\end{figure}

		To justify the mechanism, from the simulated data, we determine the azimuthal angles $\phi_\text{p}$ and $\phi_\text{m}$ from the in-plane components of spin current polarization $\mathbf{p}$ and  $\mathbf{m}$ respectively and thus we estimate the relative phase angle, $\Phi(t) = \phi_\text{p}(t)-\phi_\text{m}(t)$, between the $\mathbf{J}$  and  $\mathbf{m}$ which is depicted by the blue line in Fig. \ref{Fig3}(c). The red line in Fig. \ref{Fig3}(c) represents the transition of $\mathbf{m}$ from its initial state ($m_z=1$) to the equator. During this transition, the corresponding $\Phi(t)$ is maintained around the range of 0 to -90$^\circ$, as shown by the blue line in Fig. \ref{Fig3}(c). This behavior aligns with the expected results predicted by equation \eqref{soteng} or Fig. \ref{Eng1}(a) for the given scenario. For this time interval, the energy change rates are positive ($+\dot\epsilon_\text{SOT}$) presented by the dotted blue line (contributed by DL-SOT) and dashed blue line (contributed by FL-SOT) in the Fig. \ref{Fig4}. That is, before crossing the equator, both the DL- and FL-torques  contributed energy to  $\mathbf{m}$. After crossing the equator, $\mathbf{m}$ changes precession direction and similarly the CCP also changes the sense of rotating direction. Consequently, the $\Phi(t)$ again maintains around  $0$ to $-90^{\circ}$ by balancing the fields of  DL- and FL-torque components. In this time interval,  $\dot\epsilon_\text{SOT}$ is negative shown by the dotted blue line (DL-SOT) and dashed blue line (FL-SOT) in the Fig. \ref{Fig4}. That is, the stimulated energy emission is occurred by  $\mathbf{m}$ through the negative work done.  The total energy changing with time contributed by the full SOT is shown by solid blue line in Fig. \ref{Fig4}. Therefore,  $\mathbf{m}$ efficiently absorbs (emits) the energy before (after) crossing the equator from (to) the fields of DL- and FL-torque components, in other words, the CCP triggers the efficient energy absorption (emission) of the magnetization from (to) the FL- and DL-components of SOT before (after) crossing over the energy barrier.

\begin{figure}
			\centering
			\includegraphics[width=80 mm, height=45 mm]{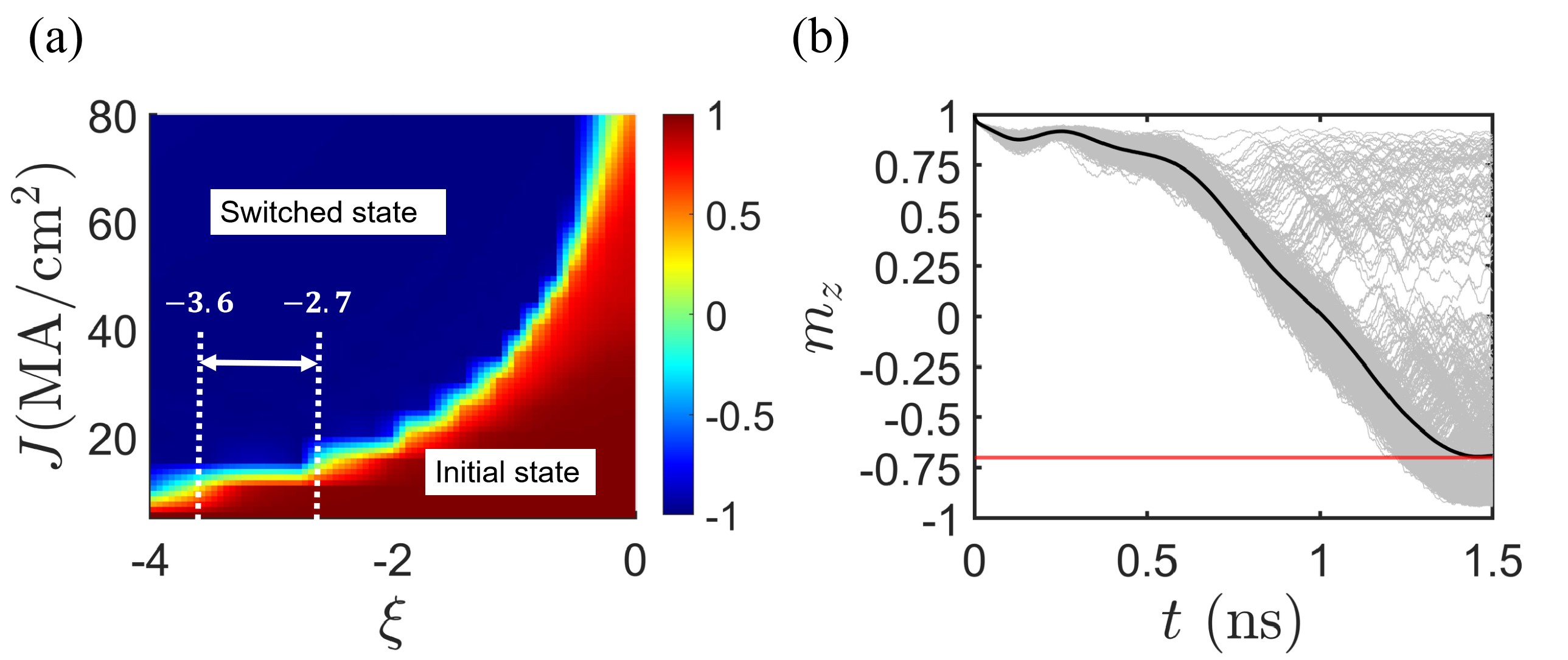}
			\caption{\label{Fig5}(a) For fixed \({f_\text{in}}\)=16.8 GHz and \(\eta=\)0.22 GHz,  the phase diagram (\(\textbf{J}-\xi\)) of magnetization switching in terms of current amplitude \(\textbf{J}\) of CCP and \(\xi\). With the switching time window of 1 ns,  the blue region indicates the switched state and red region shows the not-switched state (i.e., initial state). (b) Bold black line shows the switching of \(\mathbf{m}\) which is obtained from  average of 1000 ensembles of magnetization switching under CCP ( with the parameters \(J=18\) MA/cm\(^2\), \( f_\text{in} = 16.8\) and \(\eta=0.14\) GHz)  at 300 K for the same material parameters as before.}
		\end{figure}

		We also investigate the magnetization switching for a range of \(\xi ( \tau_{\text{FL}}/{\tau_{\text{DL}}}\))  to see how the current density of CCP changes. For fixed \({f_\text{in}}\) = 16.8 GHz, \(\eta=\) 0.22 GHz and switching time window of 1.5 ns, Fig. \ref{Fig5}(a) shows the phase diagram (\(\textbf{J}-\xi\)) of magnetization switching in terms of current amplitude \(\textbf{J}\) of CCP and \(\xi\) with the other parameters kept similar, in which the blue region indicates the switched state and red region indicates the not-switched state (i.e., initial state). It shows that for larger \(\xi\) i.e., for the FL-SOT dominated case, smaller \(\textbf{J}\) is enough to switch. Note that there is a range of  \(\xi\)(-3.6 to -2.7) (indicated by the two vertical dot lines) for which the constant current is required. This range of  \(\xi\) will provide an advantage for device application. However, it is suggested to use as larger value of  \(\xi\) as possible for an efficient CCP driven switching.

		Since the temperature is present everywhere in nature. So, to satisfy the realistic condition, it is meaningful to check the strategy whether the CCP driven magnetization reversal is valid at finite/operating temperature. Purposely, we use the stochastic Landau Lifshitz Gilbert (sLLG) equation to simulate the system with finite temperature and examine the CCP ( with the parameters \(J=18\) MA/cm\(^2\), \( f_\text{in} = 16.8\) and \(\eta=0.14\) GHz) driven magnetization reversal at 300 K for the same material parameters as before.  To find the proper estimation (using the similar approach of the studies \cite{islam2019, islam2020}), we repeat 1000 times (by changing the thermseed) of the CCP driven magnetization switching (\(m_z\)) for the same material parameters. Then take the ensemble average of \(m_z\) as shown in the Fig. \ref{Fig5}(b) and thus find that the magnetization switching (bold black line) is valid and robust even at room temperature.

\begin{figure}[t]
			\centering
			\includegraphics[width=85mm, height=60mm]{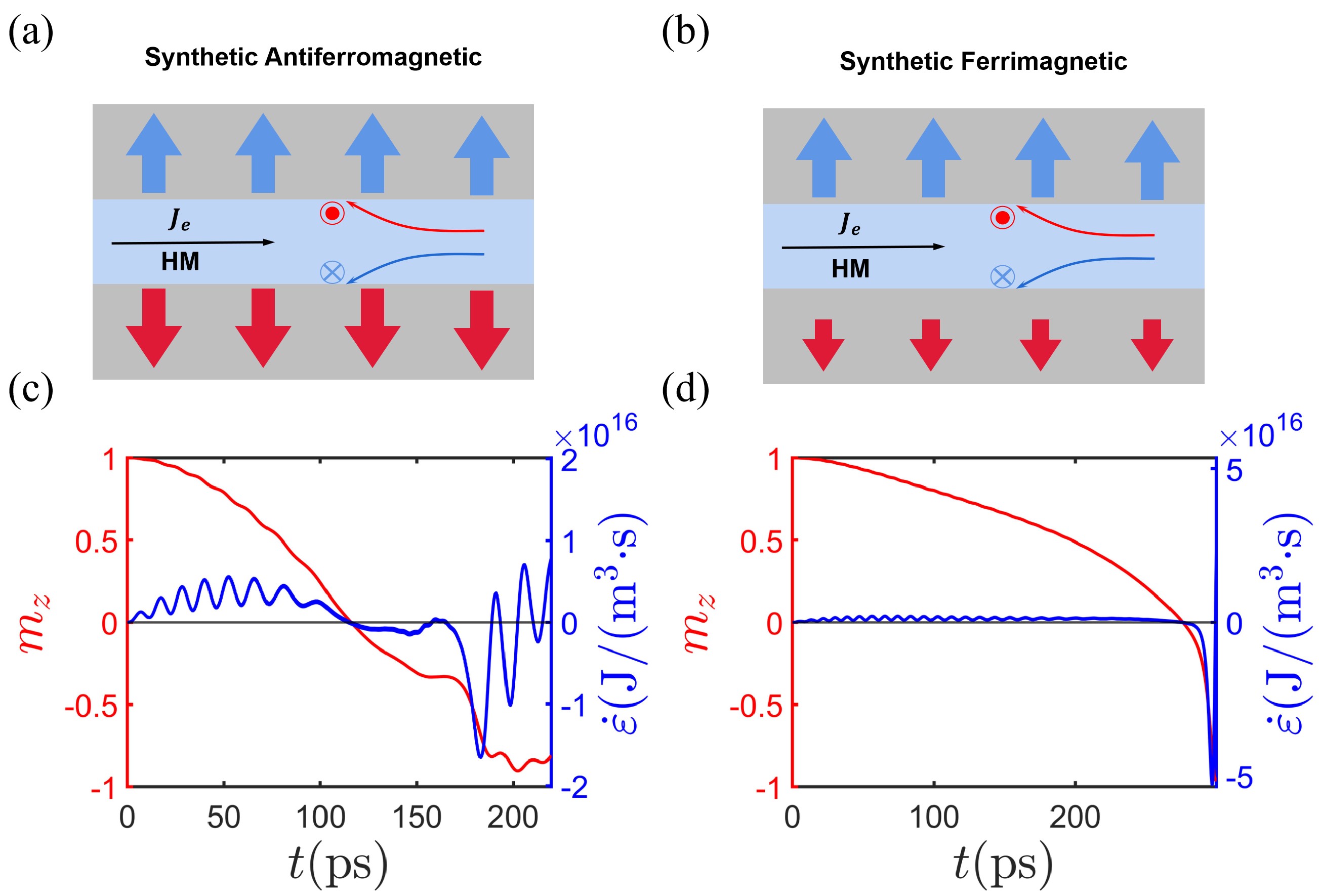}
			\caption{\label{Fig6} Sketch of perpendicular synthetic (a) Antiferromagnetic (SAF) structure  and (b) ferrimagnetic (SFi) structure. The rotating  CCP is applied in the  sandwiching a heavy metal layer. Temporal evolution of \(m_z\) (red line) of upper layer of synthetic (c) SAF  and (d) SFi structure while Blue lines (in (c) and (d)) show the energy changing rate \(\dot\epsilon_\text{SOT}\) of \(m_z\) contributed by the both DL- and FL-torques. }
		\end{figure}

		Recently, the reversal of antiferromagnetic (AFM) state draws much attention as the AFM nanostructure possesses some intriguing properties like the ultrafast spin dynamics, insensitivity against external magnetic field, zero stray field \cite{jungwirth2016antiferromagnetic, MacDonald2011,parkin1991systematic,Fukami2016}. Similar to AFM system, people also focus on magnetization reversal of the synthetic antiferromagnetic (SAF) and synthetic ferrimagnetic (SFi) nanodevice due to their promising properties of zero stray field \cite{Bandiera2010,Jenkins_2020} and  high thermal stability \cite{Yakata2009,Yoshida2013, Bran2009}. So, it is interesting to check whether the above strategy works efficiently also to reverse the magnetization of SAF and SFi nanodevices. Purposely, the schematic diagrams of SAF and SFi systems are shown in Fig. \ref{Fig6}(a) and (b) in which  two ferromagnetic layers, coupled  through the interlayer exchange interaction of  Ruderman–Kittel–Kasuya–Yoshida (RKKY) \cite{parkin1991systematic}, mediated by a sandwiched heavy metal layer. For SAF, the upper and lower layer have same dimension ($20\times 20 \times 1.2$ nm$^3$) and same material parameters as previously chosen for the FM system. We intend to apply the CCP in the sandwiched layer (which is heavy metal) and thus the transverse  spin currents, which are generated because of the Spin-Hall effect, pass through the adjacent upper and lower ferromagnetic layers as shown in Fig. \ref{Fig6}(a). So both transverse spin current components from the CCP can be utilized efficiently to switch magnetization of upper (up to down) and lower (down to up) layers in opposite directions. The demagnetization field of SAF system is zero, so the resonant frequency is obtained as $\frac{\gamma h_\text{ani}}{2\pi} =44.98$ GHz. Now we investigate the magnetization reversal by applying the CCP with $f_\text{in} = 44.98$ GHz and for different values of $J$ and $\eta$. Interestingly, we found that, the CCP, with $f_\text{in}$ 44.98 GHz, the minimal current $J=29$ MA/cm$^2$ and optimal $\eta=0.43$ GHz, is capable of driving the field-free ultrafast (around 200 picosecond) magnetization reversal of the both layers. Red line in the Fig. \ref{Fig6}(c) shows only the reversal of the upper layer while blue line shows the stimulated energy absorption/emission \(\dot\epsilon_\text{SOT}\) by/from the magnetization. Note that the behavior of the rate of change of energy absorption/emission of $\mathbf{m}$ (blue line)  before (after) crossing the energy barrier is similar to the FM system  (solid blue line in Fig. \ref{Fig4}). That is, the mechanism and explanation of the fast reversal of SAF under the CCP is analogous to the FM system.   
		   
	   	Later on, we examined the CCP driven switching of perpendicular SFi states with a small net magnetization as saturation magnetizations \(M_\text{s}\) of upper and lower layers are set as $1\times10^6$ A/m and $0.85\times10^6$ A/m respectively. In addition, other material parameters as well as the dimensions are same as the previously studied FM system. Similar to SAF, we apply the rotating CCP in the sandwich layer and thus the transverse spin currents, which are generated by the spin-Hall effect can flow in the upper and lower FM layers which are align in opposite directions as shown in the Fig. \ref{Fig6}(b). In this way, both  the spin current components (upper and lower) from the CCP can be utilized efficiently to drive the reversal of magnetic state of both layers (from up-down to down-up). Thus we interestingly found that, with the minimal current density $J=17$ MA/cm$^2$, $f_\text{in}$= 44 GHz and optimal $\eta=0.13$ GHz, the ultrafast reversals of  $\mathbf{m}$ of the both layer are achieved.  Red line in the Fig. \ref{Fig6}(d) shows the switching of \(m_z\) of a single cell from the upper layer while blue line represents the trend of the energy changing rate of the magnetization of a single cell, \(\dot\epsilon_\text{SOT}\) contributed by both DL- and FL-torque components. Note that, blue line in Fig. \ref{Fig6}(d), the trend of the rate of change of energy absorption (emission) before (after) crossing the energy barrier is similar to the FM system  (solid blue line in Fig. \ref{Fig4}). Thus the physical mechanism and explanation of the ultrafast reversal of SFi under the CCP is analogous to the FM system. However, it is mentioned that, in compare to  SAF switching, the minimal current needed for SFi switching is significantly lower.

		\section{Discussion and Conclusions}
		In this study, we investigated the parameters necessary for efficient and ultrafast switching in field-free spin-orbit torque (SOT) systems driven by a circularly polarized current pulse (CCP). We found that a wide range of initial frequencies around the ferromagnetic resonance frequency (FMR) allowed for ultrafast switching. By increasing the controlling parameter 
$\eta$, we were able to achieve ultrafast switching, as the switching time is close to the pulse duration of the CCP. Additionally, we found that increasing the current density $J$ led to more efficient energy absorption and emission. However, generating the CCP in the laboratory remains a challenge, and recent studies may provide insight into how to generate such a pulse \cite{Polley2023, Qiu2014,JKim2012,Manchon2019,Filianina2020, Ding2020}.

Our study demonstrated that utilizing both FL- and DL-torques in the CCP of 0.92ns can drive an ultrafast and efficient magnetization reversal in 0.94ns without any additional means of symmetry breaking. We found that the critical current density is strongly reduced in comparison to conventional SOT-reversal for a certain range of CCP frequencies. This field-free ultrafast reversal is driven by the efficient energy absorption and emission of the magnetization from and to the fields of FL- and DL-torques before and after crossing over the energy barrier. The pulse duration of the CCP is close to the reversal time and does not require precise adjustment. This CCP-driven switching is robust at room temperature.

We also extended this strategy to reverse the magnetic states of perpendicular SAF and SFi systems by applying the CCP in the sandwiched layer, acting as a spin source for upper and lower layers. We found that the CCP, with significantly lower current density, can induce field-free ultrafast and efficient switching of magnetic states of SAF and SFi systems with a similar physical mechanism. These findings enhance the basic understanding of SOT-reversal and provide a way to realize novel SOT-MRAM devices with FM, SAF, or SFi free layers.

		\begin{acknowledgments}
			M.T.I. acknowledges the support of National Natural Science Foundation of China (Grant No. 12350410352) and  National Key R and D Program of China (Grant No. 2021YFA1202200). X.R.W. acknowledges the support of Hong Kong RGC Grants (No. 16300522, 16300523, and 16302321).  X.S.W. acknowledges support from the Natural Science Foundation of China (NSFC) (Grants No. 11804045 and No. 12174093) and the Fundamental Research Funds for the Central Universities.
		\end{acknowledgments}
		
		\appendix
		
		\section{\label{app:A}The calculation of the rate of energy change}
		In the appendix, we present a detailed derivation of the rate of change of energy of magnetization, $\frac{d\epsilon}{dt}$, in equation (5).
		
		In this context, we treat the spin-orbit torques (SOT) as an external torque, meaning that the energy associated with the effective field generated by SOT is not accounted for in the system's energy. This enables us to concentrate solely on the inherent energy variations of the system. The effective field is defined as follows:
		\begin{equation}
			\mathbf{H}_{\text{eff}} = -\frac{1}{\mu_0 M_s}\frac{\partial \epsilon}{\partial \mathbf{m}}.
		\end{equation}
		
		By making suitable adaptations, the energy density can be formulated in the following manner:
		\begin{equation}
			\frac{\partial \epsilon}{\partial t} \equiv\dot\epsilon= -\mu_0 M_s \mathbf{H}_{\text{eff}}\cdot\frac{\partial \mathbf{m}}{\partial t}.
		\end{equation}
		
		By substituting the explicit Landau-Lifshitz-Gilbert (LLG) equation into the above equation, we derive the following expression:
		\begin{align}
			\dot\epsilon=& -\frac{\alpha \gamma \mu_0 M_S}{1+\alpha^2}[|\mathbf{H}_{\text{eff}}|^2 - (\mathbf{m} \cdot \mathbf{H}_{\text{eff}})^2]- \frac{\gamma \mu_0 M_S a_j}{1+\alpha^2} \mathbf{H}_{\text{eff}}  \notag \\
			& \cdot[(1+\xi \alpha) \mathbf{m} \times (\mathbf{m} \times \mathbf{p}) + (\xi - \alpha)(\mathbf{m} \times \mathbf{p})]\notag \\
			=&\dot\epsilon_\text{Damp}+\dot\epsilon_\text{SOT}.
		\end{align}
		
		The first term, $E_\text{Damp}$, on the right-hand side is always negative as it is proportional to $-\alpha$, and $|\mathbf{H}_{\text{eff}}|$ is always smaller than $|\mathbf{m} \cdot \mathbf{H}_{\text{eff}}|$. The second term, $E_\text{SOT}$, represents the energy variation resulting from the influence of the SOT torque, and its sign can be positive or negative. Where $\mathbf{m}$ is the magnetization direction and $\mathbf{p}$ is the spin polarization direction which is related to the electric current direction as $\mathbf{p}=\hat{J}\times\hat{z}$. $\mathbf{m}$ and $\mathbf{p}$ can be represented in spherical coordinates as:
		\begin{equation}
			\mathbf{m} = \sin\theta\cos\phi_\text{m}\,\mathbf{x} + \sin\theta\sin\phi_\text{m}\,\mathbf{y} + \cos\theta\,\mathbf{z},
		\end{equation}
		\begin{equation}
			\mathbf{p} = \cos\phi_\text{p} \mathbf{x} + \sin\phi_\text{p} \mathbf{y}.
		\end{equation}
		
		Similarly, the effective field $\mathbf{H}_{\text{eff}}$ is also a function of $\mathbf{m}$:
		\begin{align}
			\mathbf{H}_{\text{eff}} =& M_s N_{xx}\sin\theta \cos \phi_\text{m}\,\mathbf{x} + M_s N_{yy}\sin\theta \sin \phi_\text{m}\,\mathbf{y} \notag \\
			&+ (M_s N_{zz}+\frac{2K_u}{\mu_0 M_s})\cos\theta\,\mathbf{z},
		\end{align}
		
		where $N_{xx}$, $N_{yy}$, and $N_{zz}$ are the demagnetization factors in the x, y, and z directions  respectively, satisfying the condition $N_{xx}+N_{yy}+N_{zz}=1$. For our system, it is typically observed that $N_{xx}$ and $N_{yy}$ tend to approach zero. As a result, the effective field can be simplified as $\mathbf{H}_{\text{eff}} = (M_s N_{zz}+\frac{2K_u}{\mu_0 M_s})\cos\theta \mathbf{z}$. Substituting this expression into the equation yields the following:
		\begin{align}
			\dot\epsilon_\text{SOT} =& -\frac{\alpha \gamma \mu_0 M_S}{1+\alpha^2}(M_s N_{zz}+\frac{2K_u}{\mu_0 M_s})\cos\theta\sin\theta\notag \\
			&[(1+\xi\alpha)\cos\theta\cos(\phi_\text{p}-\phi_\text{m})+(\xi-\alpha)\sin(\phi_\text{p}-\phi_\text{m})].
		\end{align}
		
		Let $\Phi = \phi_\text{p}-\phi_\text{m}$, then:
		\begin{align}
			\dot\epsilon_\text{SOT} = &-\frac{\alpha \gamma \mu_0 M_S}{1+\alpha^2}(M_s N_{zz}+\frac{2K_u}{\mu_0 M_s})\cos\theta\sin\theta \notag \\
			&[(1+\xi\alpha)\cos\theta\cos\Phi+(\xi-\alpha)\sin\Phi].
		\end{align}

		\bibliography{library_JabRef}
		
	\end{document}